\documentclass[prb,preprint,superscriptaddress,nobibnotes]{revtex4}

\usepackage{times}
\usepackage[pdftex]{graphicx}
\usepackage[pdftex]{epsfig}
\usepackage{amsmath}
\usepackage{pslatex}
\usepackage{amssymb}
\usepackage{amsfonts}
\usepackage{color}
\usepackage{wrapfig}
\usepackage{eucal}
\usepackage{hhline}
\usepackage{threeparttable}
\usepackage{supertabular}
\usepackage{multirow}
\usepackage{tabularx}

\newcommand{\vecb}[1]{\mathbf{#1}}

\newcolumntype{Y}{>{\centering\arraybackslash}X}

\begin{document}

\pagenumbering{arabic}

%\title{Optical gradient force actuation of a picogram and nanometer scale photonic crystal nanocavity}
\title{A picogram and nanometer scale photonic crystal opto-mechanical cavity}
%\title{Giant optical spring in a picogram and nanometer scale photonic crystal cavity}

\author{Matt Eichenfield}
\author{Ryan Camacho}
\author{Jasper Chan}
\author{Kerry J. Vahala}
\author{Oskar Painter}
\email{opainter@caltech.edu}
\homepage{http://copilot.caltech.edu}
\affiliation{Thomas J. Watson, Sr., Laboratory of Applied Physics, California Institute of Technology, Pasadena, CA 91125}

\date{\today}

\begin{abstract}
We describe the design, fabrication, and measurement of a cavity opto-mechanical system consisting of two nanobeams of silicon nitride in the near-field of each other, forming a so-called ``zipper'' cavity.  A photonic crystal patterning is applied to the nanobeams to localize optical and mechanical energy to the same cubic-micron-scale volume.  The picrogram-scale mass of the structure, along with the strong per-photon optical gradient force, results in a giant optical spring effect.  In addition, a novel damping regime is explored in which the small heat capacity of the zipper cavity results in blue-detuned opto-mechanical damping.  
\end{abstract}

\maketitle

Recently, there has been keen interest\cite{ref:Kippenberg_Sc_review} in dynamic back-action caused by electromagnetic forces in optical~\cite{ref:Arcizet,ref:Gigan1,ref:Schliesser2,ref:Corbitt1,ref:ThompsonJD1,ref:Kippenberg4} and microwave~\cite{ref:Regal1} cavities.  Back-action cooling, for example, is being pursued as a means to achieve quantum ground-state cooling of a macro-scale mechanical oscillator.  Work in the optical domain has revolved around milli- or micro-scale structures utilizing the radiation pressure force.  By comparison, in microwave devices, low-loss superconducting structures have been used for gradient-force mediated coupling to a nanomechanical oscillator of picogram mass~\cite{ref:Regal1}.  Here we describe measurements of an optical system consisting of a pair of specially patterned nanoscale beams in which optical and mechanical energy are simultaneously localized to a cubic-micron-scale volume, and for which large per-photon optical gradient forces are realized.  The resulting scale of the per-photon force and the mass of the structure enable new cavity-optomechanical regimes to be explored, where for example, the mechanical rigidity of the structure is dominantly provided by the internal light field itself.  In addition to precision measurement and sensitive force detection\cite{ref:Stowe1}, nano-optomechanics may find application in reconfigurable and tunable photonic systems\cite{ref:Rakich1}, RF-over-optical communication\cite{ref:Hossein2}, and to generate giant optical nonlinearities for wavelength conversion and optical buffering\cite{ref:Notomi4}.       

\begin{figure}[t]
\begin{center}
\includegraphics[width=0.6\columnwidth]{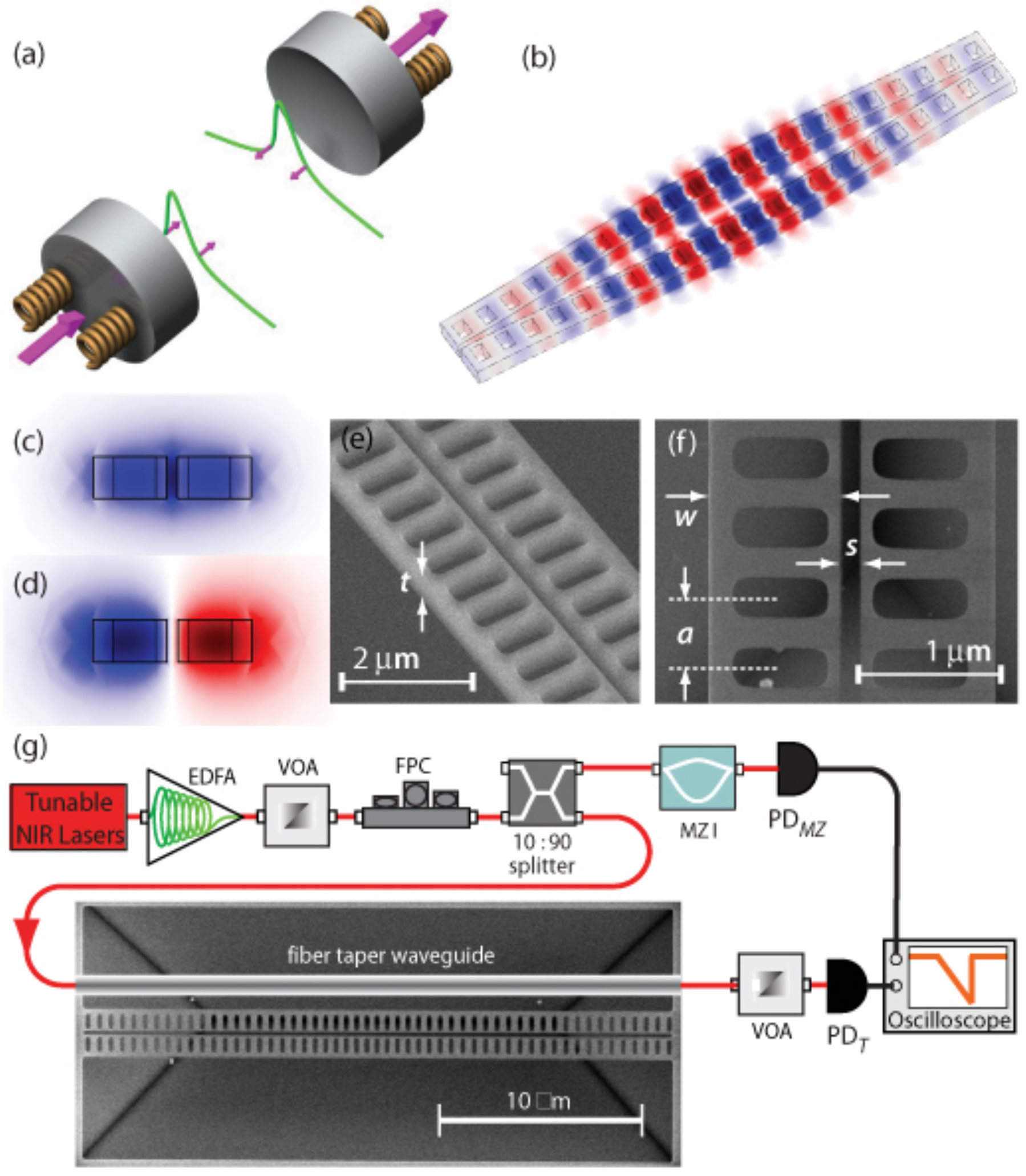}
\caption{\textbf{Zipper cavity opto-mechanical system and experimental set-up}. \textbf{a}, Fabry-Perot and \textbf{b}, photonic crystal optomecachanical systems.  Finite-element-method simulated \textbf{c}, bonded and \textbf{d}, anti-bonded supermodes of the zipper optical cavity, shown in cross-section.  \textbf{e,f}, Scanning-electron-microscope (SEM) images of a typical zipper cavity, indicating the slot width ($s$), the cantilever width ($w$), and the photonic crystal lattice constant ($a$).  \textbf{g}, Experimental set-up used to probe the optical and mechanial properties of the zipper cavity.  Acronyms are: erbium-doped fiber amplifier (EDFA), variable optical attenuator (VOA), fiber polarization controller (FPC), fiber Mach-Zender interferometer (MZI), and photodetected Mach-Zender transmission (PD$_{MZI}$) and zipper cavity transmission (PD$_{T}$).}
\label{fig:OM_struct_set_up}
\end{center}
\end{figure}

Optical forces arising from near-field effects in guided-wave structures have been proposed~\cite{ref:Povinelli1}, and recently demonstrated\cite{ref:Eichenfield1,ref:Li1}, as a means of providing large optomechanical coupling between the field being guided and the dielectric mechanical structure providing the guiding.  The resulting optical force can be viewed as an intensity gradient force much like that used to \emph{tweeze} dielectric particles or to trap cold gases of atoms\cite{ref:Ashkin1}.  In the devices studied in this work, doubly-clamped silicon nitride nanobeams are converted into optical resonant cavities through the patterning of a linear array of etched holes (Fig. \ref{fig:OM_struct_set_up}(b)).  Bringing two such cavities into the near-field of each other forms a super cavity supporting even and odd superpositions of the individual beams modes.  This ``zipper'' cavity, so-named due to its resemblence to the mechanical fastner, allows for sensitive probing and actuation of the differential motion of the beams through the internal, optical, cavity field.  

\begin{figure}[t]
\begin{center}
\includegraphics[width=0.6\columnwidth]{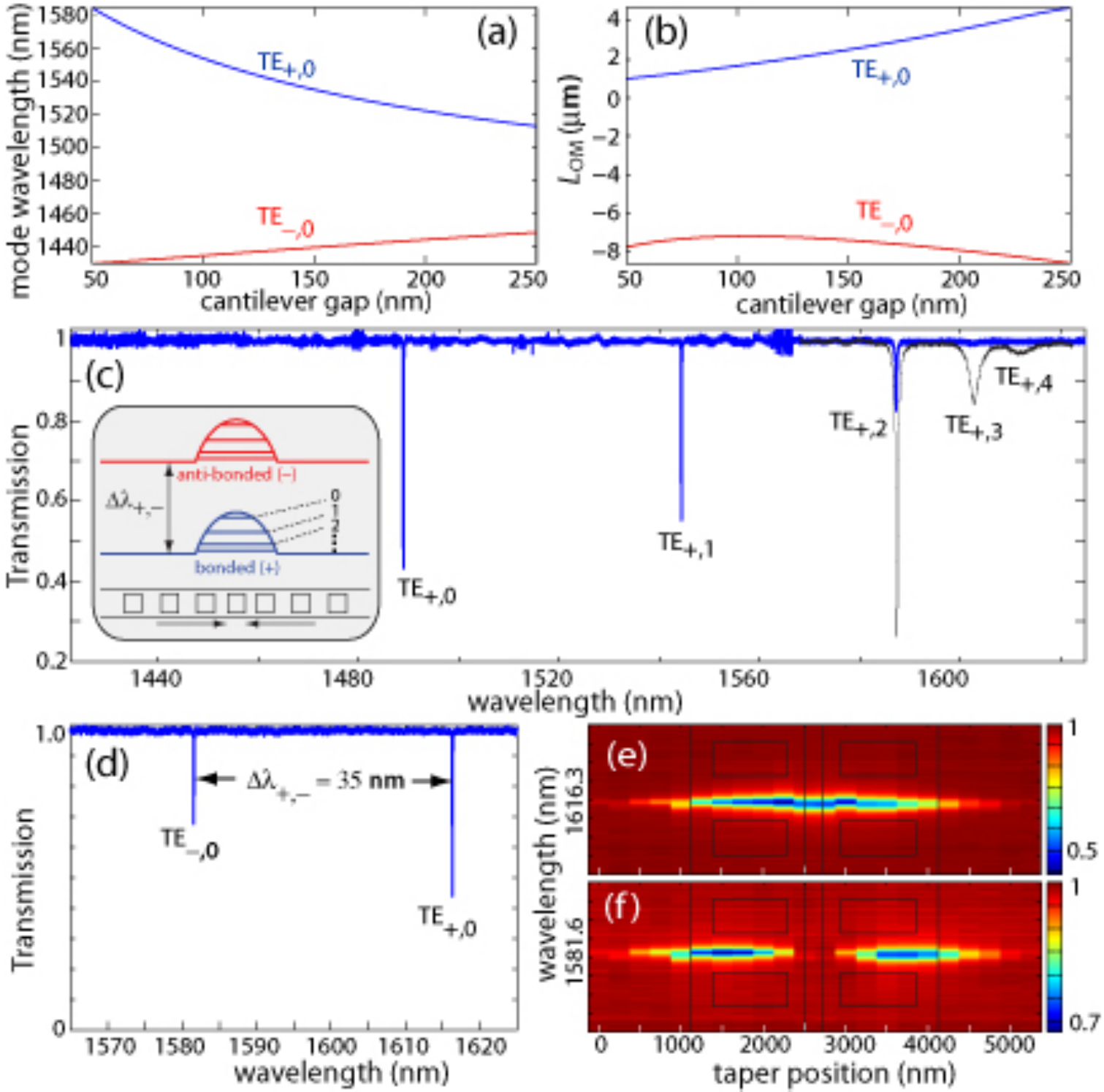}
\caption{\textbf{Optical Spectroscopy of the Zipper Cavity}.  Finite-element-method simulation of the \textbf{a}, wavelength tuning versus nanobeam slot gap and \textbf{b}, effective optomechanical coupling length parameter ($L_{\text{OM}}$) for the bonded and anti-bonded fundamental zipper cavity optical modes ($w=650$ nm).  \textbf{c}, Measured optical transmission of a zipper cavity with $w=650$ nm and $s=120$ nm showing four orders of the bonded (TE$_{+}$) resonant modes.  Inset: schematic of the graded photonic crystal lattice design and resulting bonded and anti-bonded resonance manifolds.  \textbf{d}, Measured optical transmission of a zipper cavity with a larger beam width and gap ($w=1400$ nm, $s=250$ nm) showing the bonded and anti-bonded lowest order optical resonances.  \textbf{e,f}, Optical transmission versus fiber taper lateral position for each of the bonded and anti-bonded resonant modes of \textbf{d}, indicating the even and odd parity of the modes. Black outline corresponds to position of zipper cavity.}
\label{fig:optical_measurement}
\end{center}
\end{figure}

A figure of merit for cavity-optomechanical systems is the coupling constant $g_{\text{OM}} \equiv \text{d}\omega_{c}/\text{d}x$, which represents the differential frequency shift of the cavity resonance ($\omega_{c}$) with mechanical displacement of the beams ($x$).  For the commonly studied Fabry-Perot cavity structure (Fig. \ref{fig:OM_struct_set_up}(a)), momentum transfer between the circulating light field and the mechanically-compliant end mirror(s) occurs at a rate of $2\hbar\vecb{k}_{ph}$ per round trip time, resulting in an optomechanical coupling constant that scales with the inverse of the cavity length ($L_{c}$), $g_{\text{OM}} = \omega_{c}/L_{c}$.  Similarly for whispering-gallery-mode structures, such as the recently studied microtoroid\cite{KippenbergOE}, $g_{\text{OM}}$ scales with the perimeter length through the radius of the cavity $R$, $g_{\text{OM}} = \omega_{c}/R$.  In the case of the zipper cavity the optomechanical coupling is exponentially proportional to the slot gap ($s$) between the beams, $g_{\text{OM}} = \omega_{c}/L_{\text{OM}}$ with $L_{\text{OM}} \sim w_{o} e^{\alpha s}$.  The minimum value of $L_{\text{OM}}$ is set by $w_{o}$ which is approximately equal to the beam width, while the decay constant $\alpha$ is set by the wavelength of light ($\lambda$) and the refractive index contrast of the nanobeam system.  Thus, for beam widths on the order of the wavelength of light and for a sub-wavelength slot gap, $L_{\text{OM}} \sim \lambda$, independent of the length of the nanobeams (see Fig. \ref{fig:optical_measurement}(a,b)).  This yields an optomechanical coupling more than an order of magnitude larger than can be accomplished in high-Finesse Fabry-Perot cavities\cite{ref:MillerR2} or glass microtoriod structures\cite{KippenbergOE}.  In addition, this large optomechanical coupling is realized in a versatile geometry in which motional mass and mechanical stiffness can be greatly varied, and for which the mechanical displacement energy density and optical energy density can be efficiently co-localized at optical wavelengths in the visible-NIR and for mechanical frequencies in the MHz-GHz frequency range.             

For the devices studied in this work, optically thin ($t=400$ nm) stoichiometric silicon nitride (Si$_{3}$N$_{4}$) is deposited using low-pressure-chemical-vapor-deposition on a silicon wafer in order to form the optical guiding layer and the mechanical beams.  Electron-beam lithography is used to pattern the zipper cavity consisting of beams of length $l=25-40$ $\mu$m, widths of $w=0.6$-$1.4$ $\mu$m, and with an inter-beam gap of $s=60$-$250$ nm (Fig. \ref{fig:OM_struct_set_up}(f)).  The optical cavity is created in the nanobeams by patterning holes to form a quasi-1D photonic bandgap for light (see Figs. \ref{fig:OM_struct_set_up}(e-g)).  A C$_4$F$_8$/SF$_{6}$-based plasma etch is then used to transfer the nanobeam and photonic crystal pattern into the Si$_{3}$N$_{4}$.  This is followed by a wet chemical etch of KOH which selectively etches the underlying Si subtrate and releases the patterned beams.     
      
As shown in Figure \ref{fig:OM_struct_set_up}(g), optical excitation and probing of the zipper cavity is performed using a high-efficiency optical fiber taper coupler\cite{ref:Srinivasan10} in conjunction with a bank of tunable external-cavity diode lasers.  A fiber polarization controller is used to adjust the polarization to selectively excite the transverse electric (TE) polarization modes of the zipper cavity.  The zipper optical cavity design is based upon a graded lattice concept\cite{ref:ChanJ1,ref:Srinivasan10,ref:Song} in which the lattice period is varied harmonically from the center to the ends of the nanobeam.  This results in an optical potential for photons which increases harmonically as one approaches the cavity center.  Localized modes form from photonic bands near the zone boundary with negative group velocity dispersion\cite{ref:Painter14}, with the fundamental mode of the cavity having the \emph{highest} frequency and higher-order cavity modes \emph{decreasing} in frequency (see inset of Fig. \ref{fig:optical_measurement}(c)).  Owing to the strong optical coupling between the pair of nanobeams, the photonic bands in the zipper cavity break into pairs of positive and negative parity super-mode bands.  The positive-parity superposition, designated TE$_{+}$, corresponds to a manifold of modes which have an even mode profile for the TE electric field polarization and a peak electric field intensity in the center of slot gap between the beams (Fig. \ref{fig:OM_struct_set_up}(c)).  These we term \emph{bonded} modes\cite{ref:Povinelli1}.  The negative-parity TE$_{-}$ manifold of modes (the \emph{anti-bonded} modes) have an odd parity mode profile and a node at the slot gap center (Fig. \ref{fig:OM_struct_set_up}(d)).  

By systematically varying the lattice constant of the devices, and measuring the parity of the cavity modes using the fiber taper as a near-field probe\cite{ref:Srinivasan10}, one can identify the various zipper cavity modes.  For example, for a zipper cavity with $a=640$ nm, beam width $w=650$ nm, and slot gap $s=120$ nm, the measured transmission scan across the $\lambda=1420$-$1625$ nm range is shown in Fig. \ref{fig:optical_measurement}(c).  From shortest to longest wavelength, the resonance peaks all have an even mode profile and are associated with the TE$_{+,0}$ through TE$_{+,4}$ modes of the \emph{bonded} manifold of modes.  Wavelength scans of a different zipper cavity, with larger beam width $w=1.4$ $\mu$m and slot gap $s=250$ nm, exhibits a spectrum in which the bonded and anti-bonded mode manifolds overlap (Fig. \ref{fig:optical_measurement}(d)).  The measured on-resonance transmission contrast versus lateral taper position for each of the modes is shown in Fig. \ref{fig:optical_measurement}(e), indicating their even (bonded) and odd (anti-bonded) mode character.  The optical $Q$-factor of the zipper cavity TE$_{+,0}$ mode can theoretically reach a value well above $10^6$ even in the modest refractive index afforded by the silicon nitride\cite{ref:ChanJ1,ref:McCutcheonM1}.  Experimentally we have measured zipper cavity modes with $Q$-factors in the range of $Q = 10^4-10^5$ (Finesse $\mathcal{F} \sim 10^4$), depending largely upon the fill-fraction of the air holes and their scattering of light transverse to the axis of the quasi-1D photonic bandgap.  For devices at the high end of the measured $Q$ range ($Q \sim 3\times 10^5$), we find a significant contribution to optical loss from absorption (see Appendix).

\begin{figure}[t]
\begin{center}
\includegraphics[width=0.7\columnwidth]{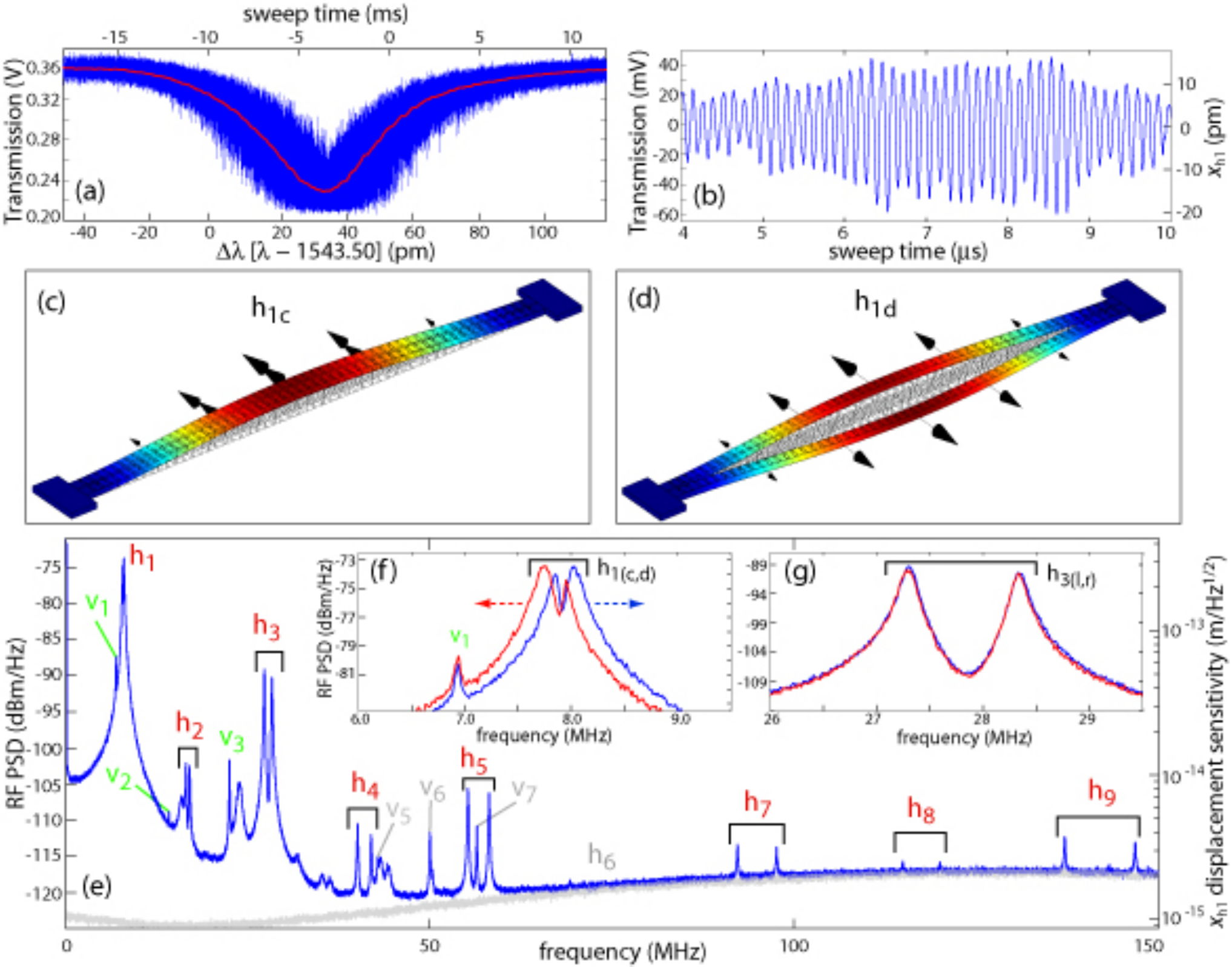}
\caption{\textbf{Mechanical Motion Transduction}.  \textbf{a}, Optical transmission through the zipper cavity of Fig. \ref{fig:optical_measurement}(c).  The red curve corresponds to the low-pass-filtered (bandwidth $10$ kHz) transmission signal showing the underlying Lorentzian-like cavity resonance. \textbf{b}, Temporal oscillations in optical transmission for fixed detuning, showing large-scale optical power oscillations of frequency $\sim 8$ MHz. FEM-modeled lowest order \textbf{c}, common and \textbf{d}, differential mechanical resonances.  Mechanical deformation and color indicates displacement amplitude, while arrows indicate direction.  \textbf{e}, Detected RF spectrum with horizontal ($\textcolor{red}{h}$) and vertical ($\textcolor{green}{v}$) cantilever modes of motion indicated.  Grey colored labels indicate either ``missing'' resonances or modes of questionable description.  Grey colored curve is the electronic detector noise floor.  Insets: zoomed-in RF spectrum of the \textbf{f}, hybridized fundamental $h_{1}$ mechanical modes and \textbf{g}, distinct left and right cantilever modes of third-order in-plane motion ($h_{3}$).  In \textbf{f} and \textbf{g} the red curves correspond to RF spectra taken for red detuning and the blue curves correspond to blue-detuned spectra.}
\label{fig:mechanical_measurement}
\end{center}
\end{figure}

Mechanical motion of the zipper cavity nanobeams is imprinted on the transmitted optical intensity through the phase modulation of the internal cavity field\cite{KippenbergOE}.  Figure \ref{fig:mechanical_measurement}(a) shows the high-temporal-resolution (blue curve) and low-pass filtered (red curve) transmitted signal as the input laser wavelength is swept across the TE$_{1,+}$ mode of the zipper cavity of Fig. \ref{fig:optical_measurement}(c) at low optical input power ($P_{i} = 12$ $\mu$W).  The zoomed-in temporal response of the transmitted intensity for a detuning on the side of the Lorentzian lineshape (Fig. \ref{fig:mechanical_measurement}(b)) shows an oscillating signal of frequency $\sim 8$ MHz and peak-to-peak amplitude of roughly a third of the transmission contrast of the resonance.  Finite-element-method (FEM) simulations (Fig. \ref{fig:mechanical_measurement}(c,d)) indicate that the lowest order in-plane common ($h_{1c}$) and differential ($h_{1d}$) mechanical modes of the pair of coupled nanobeams have frequencies of $8.19$ and $8.16$ MHz (mass, $m_{x} \approx 43$ picograms, and spring constant $k_{h1} \approx 110$ N/m; see Appendix), respectively, when accounting for $\sim 0.75$ GPa of internal tensile stress in the nitride film\cite{ref:Verbridge1}.  The corresponding mechanical amplitude of oscillation is calibrated by fitting $g_{\text{OM}}$ from the optical spring effect as discussed below and in the Appendix, yielding $L_{\text{OM}} = 1.58$ $\mu$m ($g_{\text{OM}}/2\pi = 123$ GHz/nm) and an inferred rms amplitude of motion of approximately $x_{\text{rms}} \sim 5.8$ pm.  This is in good correspondence with both the FEM-simulated optomechanical coupling constant for this device ($L_{\text{OM}} = 2.1$ $\mu$m for $s=120$ nm in Fig. \ref{fig:optical_measurement}(b)) and the expected thermal amplitude for the $h_{1d}$ mode ($\langle x_{th}^2\rangle^{1/2} = \sqrt{k_{B}T/k_{h1}} = 6.2$ pm).  

The RF spectrum of the transmitted optical intensity out to $150$ MHz is shown in Fig. \ref{fig:mechanical_measurement}(e).  Comparison to FEM mechanical simulations\cite{ref:ChanJ1} allows us to identify many of the resonances in the RF spectrum, with in-plane mechanical resonances up to $9$th-order being visible.  The strength of the corresponding spectral peaks oscillates for odd and even orders of in-plane motion, consistent with the odd-order mechanical modes having an anti-node of displacement at the center of the zipper cavity and the even-order modes having a node.  The mechanical $Q$-factor of the resonances are measured to vary between $Q_{M} \sim 50-150$, limited by gas-damping\cite{ref:Verbridge2} in the nitrogen test environment used in this work.  For the $h_{1}$ mechanical resonances (Fig. \ref{fig:mechanical_measurement}(f)) at $8$ MHz, the RF spectrum shows two other interesting features.  The first is the interference between the two resonances as evidenced by the asymmetry in each peak and the narrow central dip.  As will be detailed elsewhere, this is due to coupling between the common and differential modes of mechanical oscillation for which the common-mode motion is \emph{dark} with respect to our optical read-out method.  The second feature of interest is the slight shift of the $h_{1}$ resonance peaks to lower (higher) resonance frequencies for red (blue) laser-cavity detuning.  Both of these features are absent for the $h_{3}$ resonance peaks centered around $28$ MHz (Fig. \ref{fig:mechanical_measurement}(g)) for which the optomechanical coupling is weaker and the frequency-splitting between independent nanobeam motion is much larger than for the $h_{1}$ modes.                         

\begin{figure}[ht]
\begin{center}
\includegraphics[width=0.6\columnwidth]{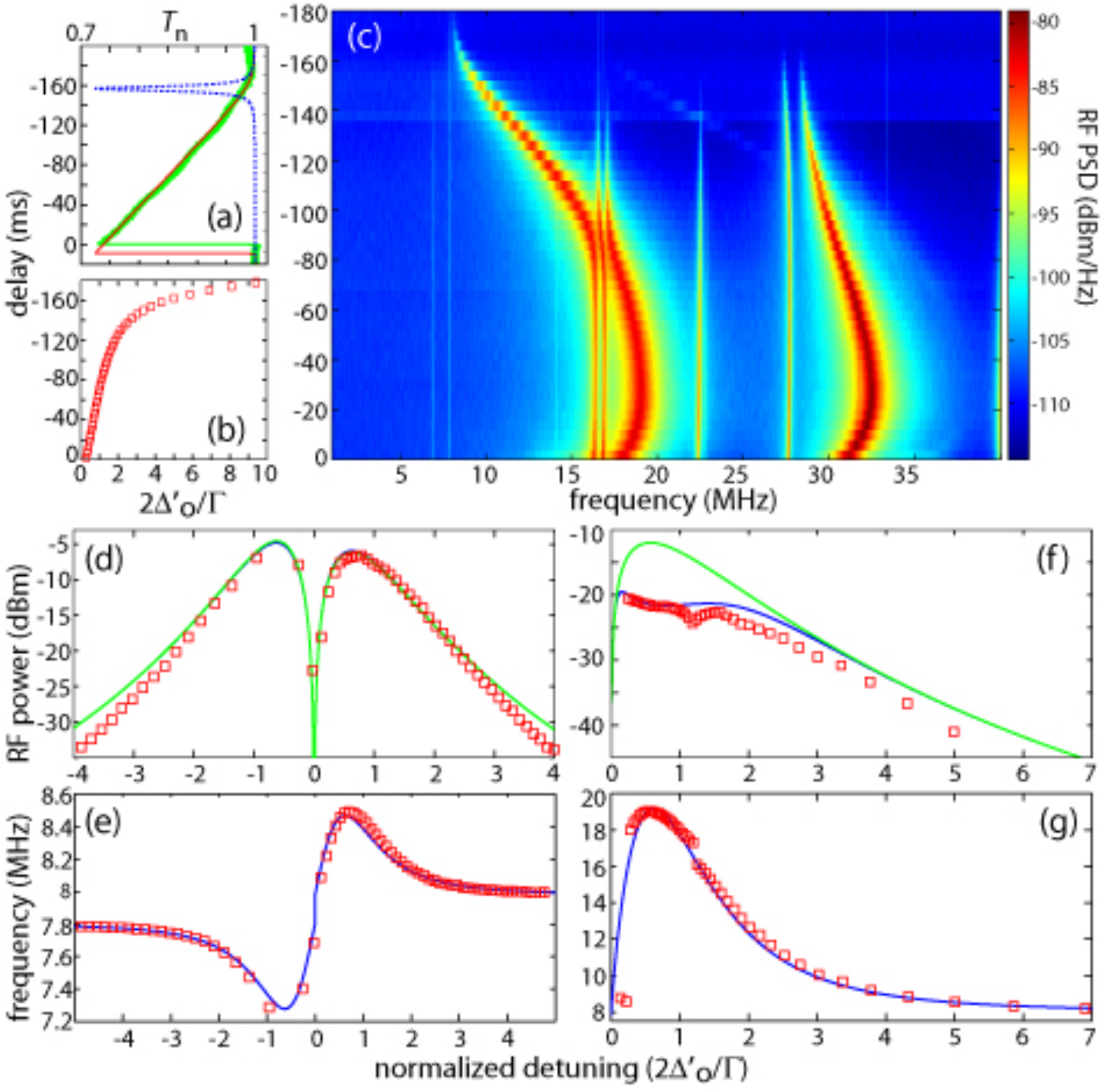}
\caption{\textbf{Optical Spring and Damping}.  \textbf{a}, Measured (green curve) and fit model (red curve) normalized optical transmission versus wavelength sweep in units of sweep time.  Dashed blue curve corresponds to low power model curve.  \textbf{b}, Conversion between sweep time delay and normalized cavity detuning, fit from model curve in (a), with zero time delay corresponding to zero laser-cavity detuning.  \textbf{c}, Intensity image of the measured RF power spectrum versus cavity detuning (time delay) of the optical transmission signal from the zipper optical cavity of Fig. \ref{fig:optical_measurement}(c) at an input optical power of $5.1$ mW (dropped power of $1.4$ mW).  Measured and modeled \textbf{d},\textbf{f} total RF power and \textbf{e},\textbf{g} resonance frequency of the $h_{1,d}$ mechanical mode, versus detuning. \textbf{d},\textbf{e} correspond to low optical input power ($P_{i}=127$ $\mu$W) while \textbf{f},\textbf{g} are for high optical input power ($P_{i}=5.1$ mW).  Blue (green) curves correspond to a model with (without) optomechanical damping.  Red squares are measured data points.}
\label{fig:spring_damping}
\end{center}
\end{figure}

The optically driven zipper cavity not only allows for sensitive mechanical displacement detection, but can also strongly modify the mechanical motion in two distinct ways.  Optical stiffening of the mechanical resonant structure\cite{ref:Corbitt1,KippenbergOE} (the so-called ``optical spring'') results from the component of optical cavity energy (and gradient force) oscillating in-phase with the mechanical motion.  On the otherhand, the finite cavity photon lifetime introduces a non adiabatic, time-delayed, component of optical force acting in-quadrature with the mechanical motion.  This velocity dependent force results in detuning-dependent amplification or damping of the mechanical motion.  A perturbative analysis shows that in the sideband unresolved limit ($\Omega_{M} \ll \Gamma$) the effective mechanical frequency ($\Omega_{M}^{\prime}$) and damping rate ($\gamma_{M}^{\prime}$) are given by the following relations (see Ref. [\onlinecite{KippenbergOE}]):

\begin{align}
(\Omega_{M}^{\prime})^2 = \Omega_{M}^2 + \left(\frac{2|a_{0}|^2g_{\text{OM}}^2}{\Delta^2\omega_{c}m_{x}}\right)\Delta_{o}^{\prime}, \label{eq:ren_Omega_M_2} \\ 
\gamma_{M}^{\prime} = \gamma_{M} - \left(\frac{2|a_{0}|^2g_{\text{OM}}^2\Gamma}{\Delta^4\omega_{c}m_{x}}\right)\Delta_{o}^{\prime}, \label{eq:ren_gamma_M_2}
\end{align}

\noindent where $\Omega_{M}$ and $\gamma_{M}$ are the bare mechanical properties of the zipper cavity, $|a_{0}|^2$ is the time-averaged stored optical cavity energy, $\Delta_{o}^{\prime} \equiv \omega_{l} - \omega_{c}$ is the laser-cavity detuning, $\Gamma$ is the waveguide-loaded optical cavity energy decay rate, and $\Delta^2 \equiv (\Delta_{o}^{\prime})^2 + \Gamma^2$.

As is shown in Fig. \ref{fig:spring_damping}, for higher optical input powers ($P_{i}=5$ mW) the internal optical cavity field provides significant stiffness to mechanical motion of modes in the zipper optomechanical cavity.  The large optomechanical coupling ($L_{\text{OM}} = 1.58$ $\mu$m) and comparitively small motional mass ($m_{x} \sim 43$ picograms) of the $h_{1d}$-zipper-cavity mechanical mode results in a giant optical spring effect\cite{ref:Corbitt1}, shifting the mechanical frequency from $8$ to $19$ MHz (Fig. \ref{fig:spring_damping}(g)).  This corresponds to an optical stiffness greater than five times that of the intrinsic mechanical stiffness of the silicon nitride cantilevers.  For the $h_{3d}$ in-plane mode, the frequency shift is smaller due to the slightly reduced optomechanical coupling factor of this mode and its larger bare frequency.  An additional feature in Fig. \ref{fig:spring_damping}(c) is the mechanical mode mixing that occurs as the optical spring tunes the $h_{1d}$ mode through other mechanical resonances.  The mode mixing is most prevelant as the $h_{1d}$ sweeps away from the $h_{1c}$ mode and through the $h_{2}$ in-plane modes near $\Omega/2\pi=16$ MHz (Fig. \ref{fig:spring_damping}(c)).  This mixing of mechanical modes is due to the highly anisotropic and motion-dependent optical stiffness and its renormalizing of the mechanical eigenmodes of the structure, and will be discussed in more detail elsewhere.  

Figures \ref{fig:spring_damping}(d-g) compare the measured integrated RF power in the $h_{1d}$ mechanical resonance line and its mechanical frequency to a nonlinear optical model of the zipper cavity system including the optical gradient force and thermo-optic tuning of the cavity (see Appendix).  At low optical power (Figures \ref{fig:spring_damping}(d,e)), a single estimate for $g_{\text{OM}}$ based upon optical FEM simulations fits both the total measured RF power (or $\langle x^2 \rangle$) and optical frequency of the $h_{1d}$ mode over a large detuning range.  At higher powers (Figures \ref{fig:spring_damping}(f,g)), the same estimated $g_{\text{OM}}$ fits the optical frequency tuning of the $h_{1d}$ mode, but severely over estimates the total RF power (green curve in Fig. \ref{fig:spring_damping}(f)) where the optomechanical interaction is strongest.  Damping of the mechanical motion is quite unexpected for blue-detuned laser excitation\cite{KippenbergOE}.  FEM numerical simulations of the zipper cavity indicate thermo-mechanical effects\cite{ref:Hohberger1,ref:Ilic1} produce a response several orders of magnitude too small to explain the observed damping; however, a theoretical analysis of the cavity dynamics including the thermo-optic effect\cite{ref:Painter_TO} shows that the severely phase-lagged and damped thermo-optic tuning of the cavity introduces a significant correction to eq. (\ref{eq:ren_gamma_M_2}).  Owing to the small heat capacity of the zipper cavity, thermo-optic tuning reverses the sign of the damping coefficient of the bare optomechanical response for blue-detuned pumping (the correction to the optical spring is found to be small, at the $10^{-4}$ level).  The numerical model including the thermo-optic correction to the spring and damping terms is shown as a blue curve in Figs. \ref{fig:spring_damping}(d,f), with the fit to the high power data now in much better agreement.  The model indicates that at a detuning of $\Delta_{0}^{\prime}\approx\Gamma/4$ (Fig. \ref{fig:spring_damping}(f)) the thermal motion of the $h_{1d}$ resonance is being damped from $x_{\text{rms}} \approx 7$ pm down to $1$ pm, at a bath temperature of $360$ K.

Beyond the giant optical spring effect afforded by the large optomechanical coupling and picogram-scale mass of the zipper cavity, thin-film photonic crystals offer a highly flexible, chip-scale architecture for coupling optical and mechanical degrees of freedom.  In the area of quantum cavity-optomechanics, significant improvements in optical $Q$ to values approaching  $5\times 10^6$ ($\mathcal{F}\sim 10^6$) can be expected with new processing technqiues\cite{ref:Borselli3,ref:ChanJ1}, which along with increased mechanical frequency ($> 100$ MHz), will push the system into the important sideband resolved limit\cite{ref:Schliesser1,ref:Marquardt1}.  Applications to optical cavity QED\cite{ref:MillerR2} also exist, where rapid cavity frequency shifting may be utilized for single-photon generation and quantum-state transfer.  Finally, by combining phononic\cite{ref:OlssonIII1} with photonic crystal concepts, simultaneous routing and localization of acoustic and optical waves can be envisioned.  Such a platform would expand both quantum and classical applications, and enable integration not possible in current optomechanical microsystems. 

\section*{Appendix}

\subsection{Optomechanical coupling, effective mass, and spring constant} For complex geometries and motional patterns, one must use a consistent definition of displacement amplitude, $x$, in determing $g_{\text{OM}}$, $m_{x}$ (motional mass), and $k_{\text{eff}}$ (effective spring constant).  In this work we use a convention in which $x(t)$ represents the amplitude of motion for a normalized mechanical eigenmode displacement field pattern:

\begin{equation}
\label{eq:motion_def}
\vecb{u}_{n}(z,t) = x_{n}(t)\frac{\vecb{f}_{n}(z)}{\sqrt{\frac{1}{l}\int_0^{l}|\vecb{f}_{n}(z)|^2\text{d}z}},
\end{equation}

\noindent where $n$ is a mode label, $l$ is the length of cantilever, and, for the simple cantilever geometry considered here, the displacement vector is only a function of position along the axis of the cantilevers ($z$).  With this definition of amplitude, the effective motional mass is simply the total mass of the two cantilevers ($m_{x}=m_{c}=43$ picograms) and the effective spring constant is defined by the usual relation $k_{\text{eff}} = m_{c}\Omega_{M}^2$, with $\Omega_{M}$ the mechanical eigenmode frequency.  The amplitude associated with zero-point motion and used in the equipartition theorem to determine the thermal excitation of the mechanical mode is then $x_{n}(t)$.  In the case of the fundamental differential mechanical mode of motion for the two cantilevers of the zipper cavity, this normalization prescription yields $\vecb{u}_{h1d}(z,t) \approx x_{h1d}(t)\left( \sin(\pi z/l)\hat{x}_{1} + \sin(\pi z/l)\hat{x}_{2}\right)$, where $\hat{x}_{1}$ and $\hat{x}_{2}$ are in-plane unit vectors associated with the two nanobeams of the zipper cavity and pointing in opposite directions away from the center of the gap between the nanobeams.  Thus, to be consistent, $g_{\text{OM}}$ for the $h_{1d}$ mode must be defined approximately as the rate of change of cavity frequency with respect to \emph{half} the change in slot gap ($g_{\text{OM}} \approx \text{d}\omega_{c}/\frac{1}{2}\text{d}\delta s$), as the amplitude $x_{h1d}(t)$ corresponds to a change in slot gap of $2 x_{h1d}(t)$ near the center of the cavity.        
    
\subsection{Optical transmission, measured RF spectra, and motional sensitivity}  RF spectra are measured by direct detection of the optical power transmitted through the zipper cavity using a $125$ MHz bandwidth photodetector (noise-equivalent-power NEP$=2.5$ pW/Hz$^{1/2}$ from $0$-$10$ MHz and $22.5$ pW/Hz$^{1/2}$ from $10$-$200$ MHz, responsivity $R = 1$ A/W, transimpedance gain $G=4\times10^4$ V/A) and a high-speed oscilloscope ($2$ Gs/s sampling rate and $1$ GHz bandwidth).  As shown in Fig. \ref{fig:OM_struct_set_up}(g), a pair of ``dueling'' calibrated optical attenuators are used before and after the zipper cavity in order to vary the input power to the cavity while keeping the detected optical power level constant.  The measured electrical noise floor is set by the circuit noise of the photodetector for the optical power levels considered in this work, corresponding to $-125$ dBm/Hz near $10$ MHz.  The motional sensitivity of the $h_{1d}$ mechanical mode is measured at $9 \times 10 ^{-16}$ m/Hz$^{1/2}$ for an optical input power of $12$ $\mu$W (corresponding to a dropped power of $3.5$ $\mu$W, and an estimated $660$ stored cavity photons).  At the power levels considered in this work, optical force noise contribution to the motional sensitivity is negligible.   

\subsection{Calibration of input power and intra-cavity photon number}  A fiber-taper optical coupling technique was used to in-couple and out-couple light from the zipper cavity.  The fiber taper, although extremely low-loss on its own ($88 \%$ transmission efficiency in this work), was put in contact with the substrate near the zipper cavity in order to mechanically anchor it during all measurements (thus avoiding power-dependent movement of the taper due to thermal and/or optical forces).  The total fiber taper transmission after mechanical anchoring of the taper to the substrate is $53 \%$.  In order to accurately determine the optical power reaching the cavity (determined by the optical loss in the taper section before the cavity) we measure the cavity response at high optical power (resulting in thermo-optic tuning of the cavity and optical bistability in the transmission response) with the input sent in one direction and then in the other of the taper.  From the asymmetry in the thermo-optic tuning in the cavity for both directions one can determine the asymmetry in the optical loss, and thus determine the optical loss before and after the zipper cavity.  Finally, this method along with calibrated measurements of the optical power at the input and output of the taper, can determine accurately the optical power reaching the zipper cavity (the \emph{input} power) and dropped by the cavity.  From calibration of the wavelength sweep using the fiber Mach-Zender interferometer one can also accurately measure the cavity linewidth and the corresponding loaded cavity $Q$.  The average stored photon number can then be determined from the dropped power and the loaded cavity $Q$.  The TE$_{1,+}$ mode is chosen to study in detail, instead of the TE$_{0,+}$ fundamental mode, due to its spectral alignment with the EDFA gain bandwidth, allowing for the higher power measurements presented in Fig. \ref{fig:spring_damping}.        

\subsection{Calibration of laser-cavity detuning}  The transduction from mechanical motion to modulated intra-cavity power, and consequently measured RF photodector spectrum, depends sensitively on the detuning point of the laser from the cavity resonance.  Accurate measurement of the laser-cavity detuning, even for large detunings ($> 5$ half-cavity-linewidths), is required to compare the theoretical model with measured data for the optical spring and damping shown in Fig. \ref{fig:spring_damping}.  Several methods exist to determine the laser-cavity detuning, including calibration of the transduced modulated photodector signal for a known mechanical or optical modulation, or simple inversion of the normalized optical transmission signal using the measured Lorentzian response of the cavity.  For the swept measurements presented in this work, we have opted to calibrate accurately the laser wavelength versus sweep time using a fiber-based Mach-Zender interferometer (FSR$=1.57$ pm at $\lambda\sim1480$ nm), and to use this to fit and convert sweep time to laser-cavity detuning by comparing with a nonlinear model of the cavity system that incorporates thermo-optic and gradient-force tuning.  The thermo-optic cavity tuning versus temperature was measured to be $14.9$ pm/K by direct measurement of the resonance wavelength shift over a $20$ K temperature range.  The optomechanical coupling constant $g_{\text{OM}}$ was estimated from both simulation, based upon an FEM model of scanning-electron-microscope (SEM) images taken of the cavity geometry, and a fit to the peak measured mechanical frequency shift.  The nonlinear cavity model, incorporating the measured thermo-optic effect and the fit $g_{\text{OM}}$, then provides an accurate conversion between wavelength and detuning from the cavity.  The above method for calibrating laser-cavity detuning is simple to employ with the swept wavelength method used in this work, and found to be much more accurate than relying on the low-pass-filtered optical transmission contrast to infer detuning (especially for large detunings where the transmission contrast is below the percent level).

\subsection{Zipper cavity optical loss}  As mentioned above, although the optical force dominates the cavity tuning at MHz frequencies, the static tuning of the cavity is still largely ($\sim 80 \%$) provided by the thermo-optic effect through optical absorption and subsequent heat generation within the zipper cavity.  Calculation of the thermal resistance of the silicon nitride zipper cavity indicates that optical absorption accounts for approximately $6 \%$ of the total optical cavity loss for the device in Fig. \ref{fig:spring_damping}(a) (an absorption-limited $Q \sim 4.8 \times 10^5$).  We attribute the optical absorption loss in the zipper cavity to surface-states\cite{ref:Borselli3} of the ``holey'' silicon nitride beams, rather than absorption in the bulk of the silicon nitride film, due to the much larger $Q$ values we have measured in less surface-sensitive microdisks formed from the same silicon nitride material.  Properly chosen chemical surface treatments should enable $Q$-factors approaching the bulk-absorption-limited value of $Q_{b} \sim 5 \times 10^6$ at $\lambda=1.5$ $\mu$m, and perhaps even higher at shorter wavelengths where optical absorption from overtones of the vibrational modes of the N-H bond is reduced.    
\section*{Acknowledgements} The authors would like to thank Qiang Lin for extensive discussions regarding the mechanical linewidth broadening of the devices studied in this work and for pointing out the origin of the mechanical resonance interference.  Funding for this work was provided by a DARPA seedling effort managed by Prof. Henryk Temkin, and by the National Science Foundation.

%\bibliography{./Optomechanics_arXiv_12_15_2008}

\end{document}